\documentclass[twocolumn,prl,tightenlines,superscriptaddress]{revtex4-2}

\usepackage{hyperref}

\usepackage{amsmath}
\usepackage{amssymb,amsfonts,latexsym}
\usepackage{bm}
\usepackage[mathcal]{euscript}
\usepackage{graphicx}
\usepackage{epsfig}
\usepackage{color}
\usepackage{xfrac}
\usepackage{mathrsfs}

\begin{document}

\title{Extreme Spontaneous Deformations of Active Crystals}

\author{Xia-qing Shi}
\affiliation{Center for Soft Condensed Matter Physics and Interdisciplinary Research, Soochow University, Suzhou 215006, China}

\author{Fu Cheng}
\affiliation{Center for Soft Condensed Matter Physics and Interdisciplinary Research, Soochow University, Suzhou 215006, China}

\author{Hugues Chat\'{e}}
\affiliation{Service de Physique de l'Etat Condens\'e, CEA, CNRS Universit\'e Paris-Saclay, CEA-Saclay, 91191 Gif-sur-Yvette, France}
\affiliation{Computational Science Research Center, Beijing 100094, China}

\date{\today}

\begin{abstract}
We demonstrate that two-dimensional crystals made of active particles can experience extremely
large spontaneous deformations without melting. 
Using particles mostly interacting via pairwise repulsive forces, we show that such active crystals
maintain long-range bond order and algebraically-decaying positional order,
but with an exponent $\eta$ not limited by the $\tfrac{1}{3}$ bound given by the (equilibrium) KTHNY theory.
We rationalize our findings using linear elastic theory and show the existence of two well-defined 
effective temperatures quantifying respectively large-scale deformations and bond-order fluctuations.
The root of these phenomena lies in the sole time-persistence of the intrinsic axes of particles, and they
should thus be observed in many different situations.
\end{abstract}

\maketitle

Two-dimensional (2D) crystalline phases are usually defined by a slow algebraic decay of positional order 
and the presence of long-range bond order. 
In thermal equilibrium, the decay exponent $\eta$ of the positional correlation function then increases
with temperature.
In many cases well-described by the celebrated Kosterlitz-Thouless-Halperin-Nelson-Young (KTHNY) theory
\cite{kosterlitz1972long,kosterlitz1973ordering,halperin1978theory,nelson1979dislocation,young1979melting}, 
melting of 2D crystals is a two-step process yielding an intermediate --usually hexatic-- phase with short-range 
positional order and only quasi-long-range bond order between the crystal and liquid phases
(see, e.g., \cite{strandburg1988two-dimensional} for a classic review). 
The first step is defined to be when thermally activated pairs of dislocations can unbind. 
This typically occurs when entropy and elastic energy for an isolated
 dislocation are balanced. KTHNY theory tells us that this must happen when $\eta$, 
 which increases with temperature, reaches a value between $\tfrac{1}{4}$ and $\tfrac{1}{3}$ 
 (for a hexagonal crystal).
This bound can be seen as putting a limit on the deformability of 2D crystals.

Crystals made of active particles are maintained out of equilibrium by the persistent injection of mechanical 
work at local time- and length-scales, and thus are not subjected to the same constraints as their equilibrium
counterparts. In particular one should not expect, on such general grounds, 
that KTHNY theory holds for the melting of 2D active crystals.
It is then surprising that, so far, most studies of the melting of simple versions of such crystals 
(without chirality nor orientational order)
have concluded that KTHNY theory still holds, or have assumed that it remains valid for deciding when 
and how melting occurs 
\cite{bialke2012crystallization,redner2013structure,singh2016universal,thutupalli2018flow-induced,cugliandolo2017phase,digregorio2018full,digregorio20192D,klamser2018thermodynamic,pasupalak2020hexatic,paliwal2020role,loewe2020solid,digregorio2022unified}.

In this Letter we show that 2D active crystals can experience extreme spontaneous deformations without melting:
they can maintain true long-range bond order and resist unbinding of dislocation pairs even as positional order correlations decay very fast, albeit still algebraically. We argue that the decay exponent $\eta$ can in fact be arbitrarily large, and rationalize our findings in terms of two well-defined 
effective temperatures quantifying respectively elastic deformations and bond-order fluctuations.

Many active crystals studied consist of spinning units or 
individual self-organized vortices forming a hexagonal lattice in 2D
\cite{riedel2005self-organized,sumino2012large,nguyen2014emergent,yeo2015collective,goto2015purely,petroff2015fast-moving,oppenheimer2019rotating,huang2020dynamical,james2021emergence,tan2022odd,oppenheimer2022hyperuniformity}.
Such chiral active crystals are investigated for their specific properties such as 
 edge modes and odd elasticity \cite{vanzuiden2016spatiotemporal,scheibner2020odd,bililign2022motile,braverman2021topological}.

Active crystals with a strong alignment of the intrinsic polarities of particles have also been considered.
Models demonstrated the possibility of traveling crystals 
\cite{gregoire2003moving,ferrante2013elasticity,ferrante2013collective,menzel2013unidirectional,menzel2013traveling,menzel2014active,ophaus2018resting,alaimo2016microscopic,weber2014defect-mediated,rana2019tuning,huang2021alignment}
in relatively small systems.
Experiments by Dauchot {\it et al.} investigated small crystals of polarly aligned particles \cite{briand2018spontaneously,vanderlinden2019interrupted}.
On the theory side, the recent work of Maitra and Ramaswamy \cite{maitra2019oriented} 
makes general predictions about oriented active solids that have not been tested so far. 

Here, we study the melting of simpler active crystals made of self-propelled particles 
subjected to pairwise repulsive forces and to rotational noise, with no or weak alignment of their polarities.
Previous works have concluded or assumed that KTHNY theory remains relevant (see in particular \cite{digregorio20192D,klamser2018thermodynamic,pasupalak2020hexatic,paliwal2020role}).
They often rely on the numerical estimation of the decay of positional order correlation functions,
a difficult task given that only relatively small system sizes are typically considered and that
the functions usually considered are intrinsically oscillatory.
Below we circumvent this problem, which allows to obtain much clearer results.

\begin{figure*}[tbp!]
    \includegraphics[width=\textwidth]{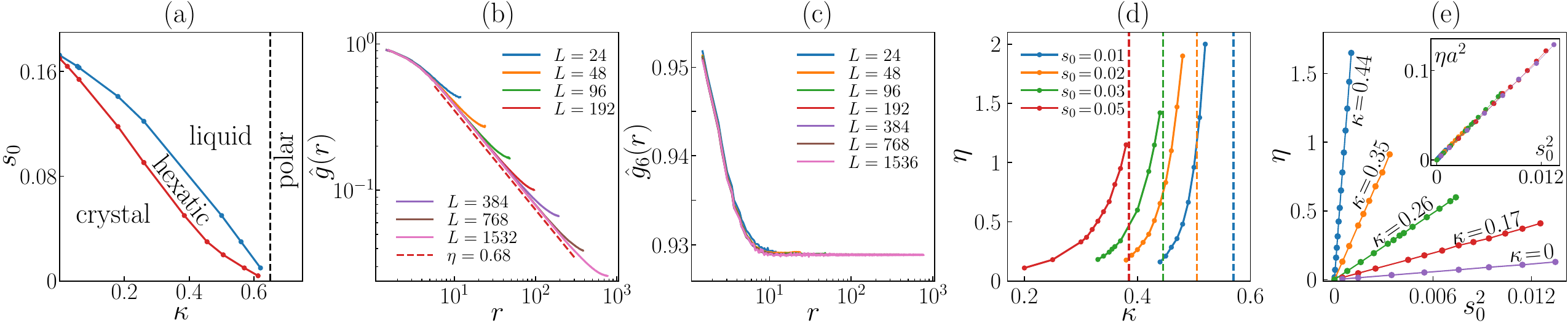}
    \caption{Model defined by Eqs.~(\ref{eqr},\ref{eqtheta}) with $d_0=D_r=1$, $\mu_r=1.5$.
    (a) phase diagram in the $(\kappa,s_0)$ plane (details about its elaboration in \cite{SUPP}).
    (b) positional order correlation function $\hat{g}(r)$ in the crystal phase ($\kappa=0.35$, $s_0=0.05$, various system sizes).
    (c) same as (b) but for the bond order correlation function $\hat{g}_6(r)$.
    (d) decay exponent $\eta$, extracted from plots similar to (b), as a function of $\kappa$ at various $s_0$ values; the dashed vertical lines indicate our estimates of the melting transition.
    (e) linear variation of $\eta$ with $s_0^2$ at various $\kappa$ values; 
    inset: same data rescaled by $a^2$, where $a$ was extracted from noise spectra such as in Fig.~\ref{fig2}(a)
    (see main text).
}
\label{fig1}
\end{figure*}

We first consider $N$ overdamped active Brownian particles interacting via soft harmonic pairwise forces 
and local ferromagnetic alignment
in 2D domains with periodic boundary conditions. Their positions ${\bf r}_i$ and polarities $\theta_i$ obey:
\begin{align}
\dot{\bf r}_i &=  s_0 \,{\bf e}(\theta_i) + \mu_r \sum_{j\sim i} (d_0-r_{ij}) \,{\bf e}_{ij} \label{eqr}\\
\dot\theta_i &= \kappa \sum_{j\sim i} \sin(\theta_j -\theta_i) + \sqrt{2D_r}\xi_i(t) \label{eqtheta}
\end{align}
where ${\bf e}(\theta_i)$ is the unit vector along $\theta_i$, $r_{ij}=|{\bf r}_j-{\bf r}_i|$,
${\bf e}_{ij}$ is the unit vector pointing from $j$ to $i$,
the sums are taken over all $j$ particles within distance $d_0$ of $i$, 
and $\xi_i(t)$ is a zero-mean Gaussian white noise 
with $\langle\xi_i(t)\xi_j(t^\prime)\rangle=\delta_{ij}\delta(t-t^\prime)$.
Numerical details are given in \cite{SUPP}.

We fix $d_0=1$, $\mu_r=1.5$ without loss of generality, work mostly with the rotational noise strength $D_r=1$, 
and vary the two remaining parameters, the self-propulsion speed $s_0$ and the alignment strength $\kappa$.
To set the stage, we first show the phase diagram of our model in this $(s_0,\kappa)$ plane (Fig.~\ref{fig1}(a),
methodological details in \cite{SUPP}). 
Strong enough alignment ($\kappa\gtrsim 0.65$) yields global polar order ($\Psi=|\langle e^{i\theta_j}\rangle_j | > 0$). In this work, we focus on the complementary region where repulsive interactions dominate 
alignment, and thus $\Psi\approx 0$, with polarity correlations decaying exponentially and very fast.
The crystal phase of main interest here is found at small $s_0$ or $\kappa$ values. 
A liquid phase is found at large $s_0$ or $\kappa$ values. 
These two phases are separated by a region of hexatic order,
in line with the KTHNY two-step scenario. 

Our primary interest being the stability of crystalline arrangements, we consider
initially perfect hexagonal configurations with lattice step $\ell_0=\tfrac{\sqrt{3}}{2}$ 
yielding a mean number density $\rho\simeq1.54$. Below, $L$ is the linear size in $y$ of approximately 
square domains (see \cite{SUPP} for details).

Positional order in the crystal phase is estimated via the large-scale behavior of the two-point correlation function
\cite{engel2013hard}
\begin{equation}
\label{eq:g}
\hat{g}(r) = \left\langle \frac{\sum_{j\ne k}\delta(r-|\hat{\bf r}_j - \hat{\bf r}_k|) e^{i\hat{\bf G}\cdot[{\bf u}_j - {\bf u}_k]}}{\sum_{j\ne k}\delta(r-|\hat{\bf r}_j - \hat{\bf r}_k|)} \right\rangle_t
\end{equation}
where $\hat{\bf r}_i$ is the position of particle $i$ on the initial perfect lattice, 
${\bf u}_i(t)={\bf r}_i(t)-\hat{\bf r}_i$ its displacement vector,
and $\hat{\bf G}$ is one of the reciprocal vectors of the perfect lattice.

The correlation function $\hat{g}_6(r)$ of orientational bond order
(which is hexatic for our triangular lattice crystals)
can be defined similarly, replacing the exponential
by $\psi_6^*(j)\psi_6(k)$ with the local order 
$\psi_6(j)=\langle e^{i6\theta_{jj'}} \rangle_{j'\sim j}$ 
where $j'$ denotes the Voronoi neighbors of particle $j$ 
and $\theta_{jj'}$ is the orientation of ${\bf r}_{j'}-{\bf r}_j$.

The  $\hat{g}(r)$ and $\hat{g}_6(r)$ correlation functions
are suited to configurations with at most bounded pairs of dislocations. 
They have the great advantage of not being oscillatory, so that their asymptotic behavior can be easily measured.

As expected, quasi-long-range positional order is observed in the crystal region, $\hat{g}(r) \sim r^{-\eta}$,
as shown in Fig.~\ref{fig1}(b), where results obtained at increasing system sizes are 
plotted~\footnote{Measurements 
performed using the more traditional, but oscillatory correlation function
\unexpanded{$g(r) = \langle \sum_{j\ne k}\delta(r-|{\bf r}_j - {\bf r}_k|) e^{i{\bf G}\cdot[{\bf r}_j - {\bf r}_k]}/\sum_{j\ne k}\delta(r-|{\bf r}_j - {\bf r}_k|) \rangle_t$}, shown in \cite{SUPP}, give essentially the same results, albeit 
with not as clean an algebraic decay as observed with $\hat{g}(r)$.}.
Strikingly, the value $\eta\simeq 0.68$ measured in this case is much larger than $\tfrac{1}{3}$,
the bound set by KTHNY theory, yet true long-range bond order is present (Fig.~\ref{fig1}(c)),
and no unbound dislocation (and hardly any bound dislocation pair) is found.
Repeating these measurements at different values of alignment strength $\kappa$ and
self-propulsion force $s_0$, we find that $\eta$ increases with both $\kappa$ and $s_0$ 
(Figs.~\ref{fig1}(d,e))
Interestingly, we observe that $\eta\propto s_0^2$ (at constant $\kappa$). 

We now present general arguments advocating for the possibility to observe $\eta$ values
larger than $\tfrac{1}{3}$ in active crystals.
In thermal equilibrium, linear elastic theory \cite{nelson1979dislocation,kardar_2007} leads to 
\begin{equation}
\eta = k_B T |\hat{\bf G}|^2 \frac{3\mu+\lambda}{4\pi\mu(2\mu+\lambda)}
\label{eq:eta}
\end{equation}
where $\mu$ and $\lambda$ are the Lam\'e elastic constants. 
Melting occurs roughly when entropy and elastic energy for creating a dislocation are balanced. This statement,
made precise by KTHNY \cite{strandburg1988two-dimensional}, yields the melting temperature
\begin{equation}
k_B T_m = \ell_0^2 \frac{\mu(\mu+\lambda)}{4\pi(2\mu+\lambda)} \;.
\label{eq:Tm}
\end{equation}
Combining these two equations by eliminating $k_BT$, and using $|\hat{\bf G}|^2 \ell_0^2=\tfrac{16}{3}\pi^2$ 
for triangular lattices, yields $\eta=\frac{(\mu+\lambda)(3\mu+\lambda)}{3(2\mu+\lambda)^2}$ at melting, 
an expression bounded from above by $\tfrac{1}{3}$.

\begin{figure*}[t!]
    \includegraphics[width=\linewidth]{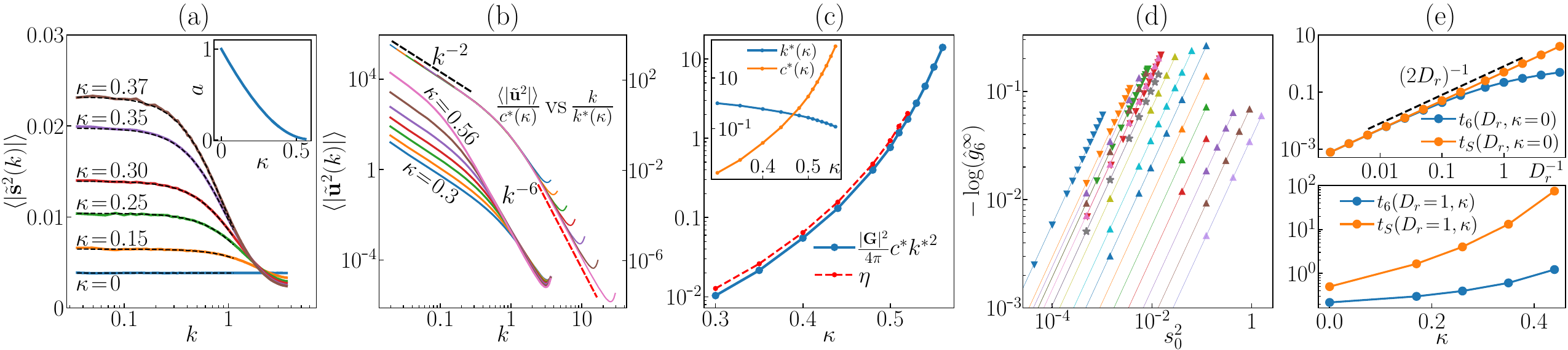}
    \caption{
    (a) Spatial spectra of ${\bf s}$ for various $\kappa$ values. 
    Black dashed curves are fits by the function $\langle|\tilde{{\bf s}}^2(k)|^2\rangle=s_0^2D_r\rho/(a+bk^2)$ 
    obtained by solving Eq.~\eqref{eq-s} \cite{SUPP}, from which we extract estimates of $a$. 
    Inset: variation of $a$ with $\kappa$. 
    (b) Spatial spectra $\langle|{\bf u}(k)|^2\rangle$ for 
    $\kappa=0.3$, 0.35, 0.4, 0.44, 0.48, 0.52, 0.56 from bottom up (lower curves) and their
    collapse $\langle|{\bf u}(k)|^2\rangle/c^*(\kappa)$ vs $k/k^*(\kappa)$
    obtained by choosing appropriate scaling parameters $c^*(\kappa)$ and  $k^*(\kappa)$. 
    (c) variation with $\kappa$ of $|\hat{\bf{G}}|^2c^*{k^*}^2/4\pi$ ($|\hat{\bf{G}}|^2=4\pi^2$)
    and $\eta$ measured as in Fig.~\ref{fig1} from the decay of $\hat{g}(r)$;
    inset: $c^*(\kappa)$ and  $k^*(\kappa)$ values used to collapse spectra in (b).
    (d)  $-\log\hat{g}_6|_{r\rightarrow\infty}$ vs $s_0^2$ for $D_r=1$ and $\kappa=0$ (grey stars),
    $\kappa=0.17$, 0.26, 0.35, 0.44 (down triangles from right to left), and
    for various $D_r=\tfrac{1}{8}, \tfrac{1}{4}, \tfrac{1}{2}, 2, 5, 10, 20, 40, 80, 160, 320, 640$ at $\kappa=0$
    (up triangles)
    (solid lines are --near perfect-- fits to linear functions of $s_0^2$)
    (e) variation with $D_r$ (top) and $\kappa$ (bottom) of $s_0^2$-rescaled temperatures $t_6$ and $t_S$ extracted from fits shown in (d) and from corresponding estimates of $\eta$ (see text).
    ($L=384$, $s_0=0.05$ in (a), $s_0=0.01$ in (b-e))}
\label{fig2}
\end{figure*}

Of course, things can be very different out of equilibrium. 
We argue here and show later that one can, in a sense, still invoke the two equilibrium relations \eqref{eq:eta} and \eqref{eq:Tm} but at two different noise levels or effective temperatures: 
$T_S$, involved in \eqref{eq:eta}, characterizes the large scale elastic deformations of the crystal structure, 
and $T_D$, involved in \eqref{eq:Tm}, drives dislocation motion. 
Given that the weak polar alignment of persistent polarities smoothes out local irregularities, 
one can expect $T_D < T_S$, which, when \eqref{eq:eta} and \eqref{eq:Tm} are combined, yields
$\eta$ values at melting larger than $\tfrac{1}{3}$~\footnote{Note that $\mu$ and $\lambda$ in principle depend on temperature, something we neglect in this rough argument.}.

These ideas can be made more concrete by considering the polarity field 
${\bf s}=s_0{\bf e}(\theta)$. Our observation that $\eta\propto s_0^2$ suggests that
{\bf s} can be taken as an effective space- and time-correlated noise.
We have calculated the spatial power spectrum of ${\bf s}$ (Fig.~\ref{fig2}(a)).
The plateaus at small wavenumber define an effective temperature corresponding to $T_S$,
while a lower noise level $T_D$ can be attributed to the $k>1$ (near lattice spacing) behavior
\footnote{For $\kappa=0$ (no alignment), unsurprisingly, the spatial spectrum is flat. 
This does not necessarily mean that $T_S = T_D$, since the ${\bf s}$ field is still time-correlated}.
These spectra 
can be superimposed by dividing them by $s_0^2$ (not shown), 
whereas the low-$k$ plateau values corresponding to $T_S$ increase
with $\kappa$ (inset of Fig.~\ref{fig2}(a)). 
Note that this is consistent with our observation that $\eta$ increases with $s_0^2$ (linearly) 
and with $\kappa$ (Fig.~\ref{fig1}(d,e)).

Most of our findings can be accounted for within linear elastic theory. 
The equation governing the displacement field ${\bf u}$ can be expressed as \cite{chaikin_lubensky_1995,huang2021alignment}:
\begin{equation}
\partial_t {\bf u} = (\lambda+\mu) \nabla (\nabla\cdot{\bf u}) +\mu\nabla^2 {\bf u} + {\bf s}
\label{eq-u}
\end{equation}
In equilibrium, {\bf s} is a white noise defining a thermal temperature $k_bT$,
but in our active crystal one can expect~\footnote{See, e.g., discussion of the time-dependent Ginzburg-Landau equation in Kardar's book \protect\cite{kardar_2007}.}
\begin{equation}
\partial_t {\bf s} = -a {\bf s} + b \nabla^2 {\bf s} + {\bm \sigma}
\label{eq-s}
\end{equation} 
with $\bm\sigma$ a white noise with 
$\langle \sigma_\alpha({\bf r},t)\sigma_\beta({\bf r}^\prime,t^\prime)\rangle$ $=s_0^2D_r\rho\,\delta_{\alpha\beta}\delta({\bf r}-{\bf r}^{\prime})\delta({t}-{t}^{\prime})$ 
(as can be obtained from coarse-graining Eqs.~(\ref{eqr},\ref{eqtheta})). Coefficients $a$ and $b$ are positive:
$-a{\bf s}$ is a damping term accounting for the short range correlations of polarities, and the Laplacian arises from
the weak aligning interactions. 

The spatial spectrum of {\bf u} can be calculated from the linear Eqs.~(\ref{eq-u},\ref{eq-s}) \cite{SUPP}. 
Its small wavenumber limit reads
\begin{equation}
\lim_{k\to0} \langle|\tilde{\bf u}(k)|^2\rangle = \frac{\rho(\lambda+3\mu)}{\mu(\lambda+2\mu)}\frac{s_0^2D_r}{2a^2k^2} \;,
\label{eq-uk}
\end{equation}
which, compared to the equilibrium case, yields an effective temperature 
$T_S=\tfrac{1}{2}s_0^2D_r/a^2$~\footnote{{In the $\kappa=0$ non-aligning limit, where the {\bf s} spectrum is flat, one has $a=D_r$, and thus $T_S=\tfrac{1}{2}s_0^2/D_r$, as expected for active Brownian particles, see e.g. \cite{cates2015motility}.}}.
Good agreement is found between linear theory and the numerical estimates of 
$\langle|\tilde{\bf u}(k)|^2\rangle$:
at fixed $s_0$, varying $\kappa$, rescaling wavenumbers by $k^*(\kappa)$ and 
spectra by a coefficient $c^*(\kappa)$ yields an excellent collapse which
reveals a $1/k^2$ behavior at low $k$ followed by a $1/k^6$ decay at high $k$ 
(Fig.~\ref{fig2}(b), and calculation in \cite{SUPP}).
The product $c^*(\kappa){k^*}^2(\kappa)$ provides an estimate of the prefactor of $1/k^2$ in Eq.~\eqref{eq-uk},
and thus, using Eq.~\eqref{eq:eta}, of the decay exponent $\eta$, which can be re-expressed as
$\eta=\tfrac{1}{4\pi}|\hat{\bf G}|^2{c^*}{{k^*}^2}$.
As expected, these estimates of $\eta$ 
match very well those obtained by directly calculating the decay of $\hat{g}(r)$ (Fig.~\ref{fig2}(c)). 

\begin{figure}[t!]
    \includegraphics[width=\linewidth]{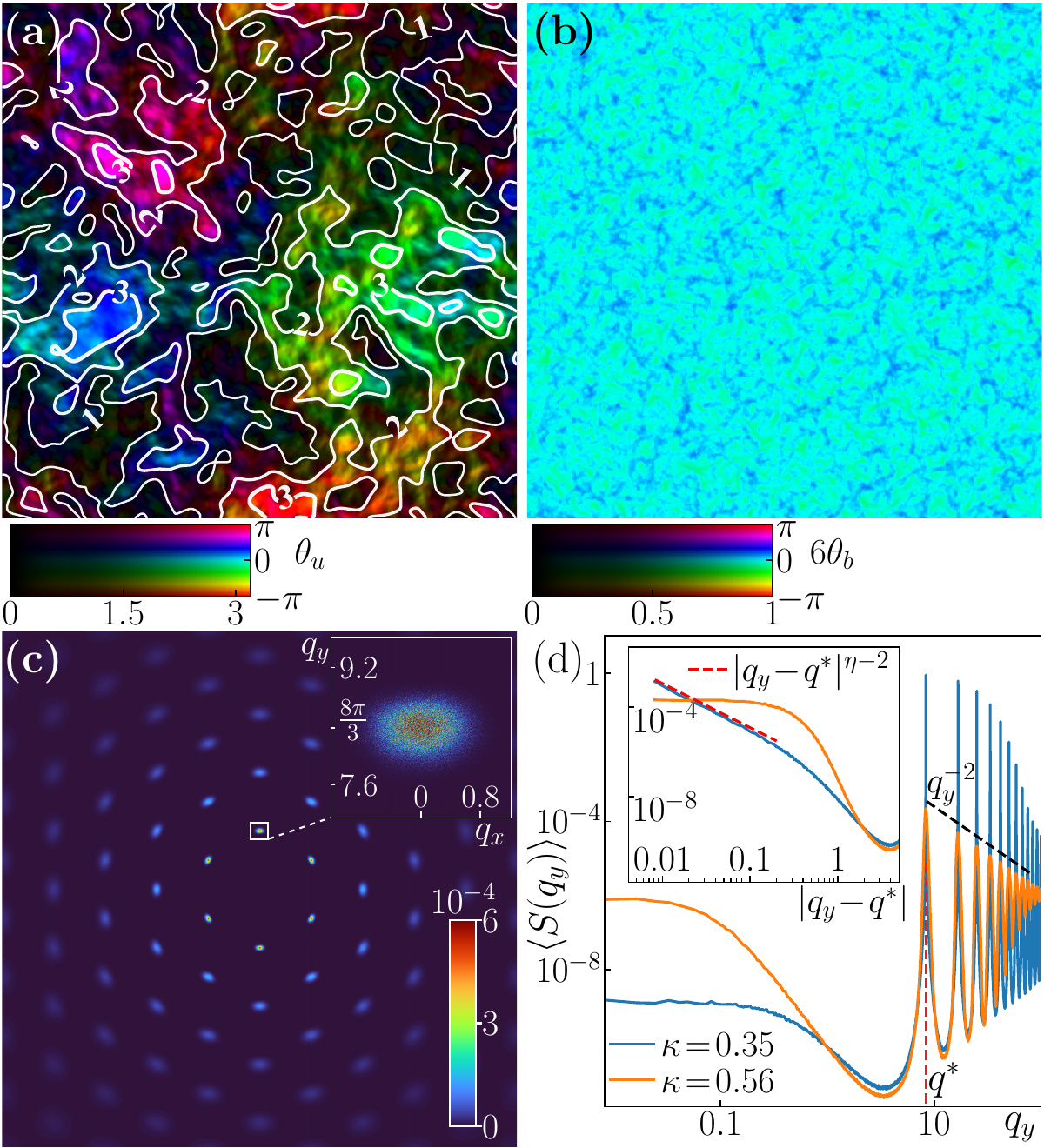}
        \caption{Highly distorted configuration in the crystal phase 
        for which the effective decay exponent $\eta\simeq 14$
        ($\kappa =0.56$, $s_0=0.01$, $L=1536$). 
        (a) Displacement field ${\bf u}$ (color is orientation, brightness is magnitude) with superimposed contour lines
        drawn for integer values of $|{\bf u}|$ in units of lattice spacing. 
        (b) Hexatic bond order field $\psi_{6}$ (color scheme as in (a)).
        (c) Instantaneous static structure factor $S({\bf q})=\rho^*(\bf q)\rho(\bf q)$ where 
        $\rho({\bf q})=\langle{\rm e}^{i{\bf q}\cdot {\bf r}_j}\rangle_j$ (color in linear scale);
        the inset is a zoom on the region of the first quasi-Bragg peak.
        (d) Time-averaged static structure factor $\langle S(q_y)\rangle_t$ measured at $q_x=0$
        superimposed on that obtained for $\kappa=0.35$ for which $\eta\simeq0.027$.
        Inset: zoom on the first peak located at $q=q^*$ showing an algebraic divergence
         $|q_y-q^*|^{\eta-2}$  only for $\kappa=0.35$.}
\label{fig3}
\end{figure}

Remarkably, the spectra-based $\eta$ values extend even further than the direct ones, 
reaching $\eta\simeq 20$ for the case presented in Fig.~\ref{fig2}(c)!
The direct evaluation of $\eta$ breaks down when $\eta\to 2$ since then the very notion of 
Bragg peak with power-law divergence disappears \cite{kardar_2007}. 
Nevertheless, even though distortions of the crystal 
exceed the lattice spacing, the deformation field {\bf u} behaves smoothly, locally preserving the crystalline order while hexatic bond order is clearly long-range (Fig.~\ref{fig3}(a,b)).
As expected, the structure factor presents a six-fold symmetry in Fourier space (testifying of crystal order), 
but it consists of smooth, non diverging, noisy extended peaks 
(Fig.~\ref{fig3}(c)). The central region flattens as $|{\bf q}|\!\rightarrow\! 0$
and the further peaks in the high $q$ region show a $q^{-2}$ scaling. 
All this is qualitatively different from the behavior
at smaller $\kappa$ values where $\eta$ is smaller than 2 and peaks with a power-law divergence 
$|q_y-q^*|^{\eta-2}$ are observed (Fig.~\ref{fig3}(d)).
Thus, estimates of $\eta$ obtained via $\langle|\tilde{\bf u}(k)|^2\rangle$ remain well defined. 
Contrary to direct measurements, they can be extended almost up to the crystal-hexatic transition.

We now consider the bond order properties of our active crystals.
At fixed parameter values, 
one can estimate the asymptotic non-zero value 
$\hat{g}_6^\infty=\lim_{L,r\to\infty} \hat{g}_6(r)$
from data such as those in Fig.~\ref{fig1}(c). 
In equilibrium, linear theory predicts that $\hat{g}_6^\infty$ decreases exponentially with temperature \cite{kardar_2007}. 
Here we do observe that $-\log(\hat{g}_6^\infty)\sim s_0^2$ for a variety of $D_r$ and $\kappa$ values (Fig.~\ref{fig2}(d)). This suggests the existence of a ``bond-order temperature'' $T_6=t_6(D_r,\kappa) s_0^2$ with
the rescaled temperature $t_6(D_r,\kappa)$ proportional to the prefactor of $s_0^2$ 
for the lines of Fig.~\ref{fig2}(d).
To be a meaningful temperature, though, the absolute value of $T_6$ must be adjusted so that it coincides with $T_S$ in the equilibrium $D_r\to\infty$ limit where the system is subjected to a white noise. 
Since $T_S=\tfrac{D_r}{2a^2} s_0^2$, we define a rescaled temperature 
$t_S(D_r,\kappa)=T_S/s_0^2=D_r/2a^2$, and globally adjust $t_6$ 
so that it coincides with $t_S$ in the $D_r\to\infty$ limit.
Our data then show how $t_6$ departs from $t_S$ as $D_r$ decreases, 
even in the absence of alignment ($\kappa=0$). Similarly, $t_6<t_S$,
and increasingly so, when $\kappa$ is increased (Fig.~\ref{fig2}(e)). 
That $T_6$ can easily be much smaller than $T_S$ indicates that fluctuations of bond order can remain rather gentle
while positional order ones are strong. Since bond disorder helps trigger the unbinding of dislocation pairs, 
one can argue that $T_6$ is closely related to $T_D$, confirming the scenario introduced earlier.

\begin{figure}[t!]
    \includegraphics[width=\linewidth]{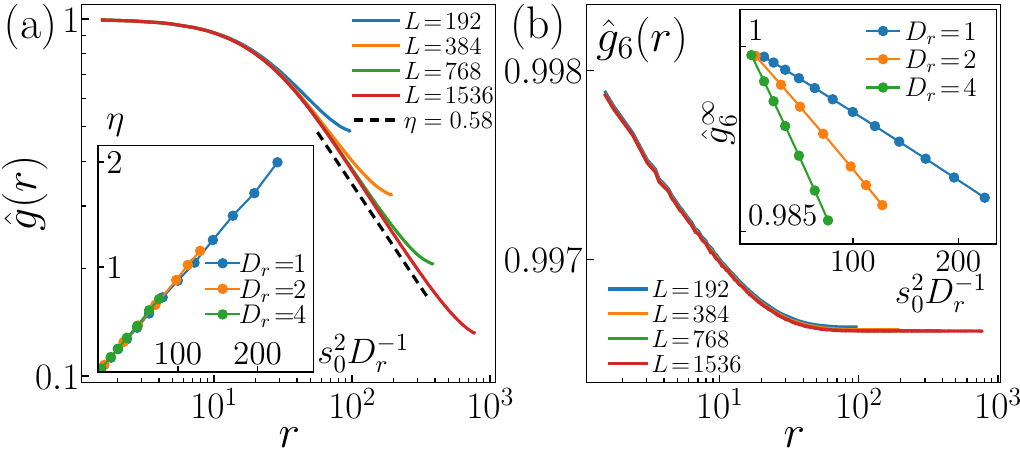}
    \caption{Crystal phase of active Brownian particles interacting via a repulsive WCA potential 
    (initial conditions as in other figures, $\rho\sigma^2=1.25$, $\epsilon=\mu_r=1$). 
    (a) 
    $\hat{g}(r)$ in the crystal phase ($s_0=8$, various system sizes), showing a decay exponent $\eta\simeq0.58$;
    inset: $\eta$ vs $s_0^2/D_r$ for various $D_r$ and $s_0$ values ($L=768$).
    (b) same as (a) but for the bond order correlation function $\hat{g}_6(r)$; 
    inset: $\hat{g}_6^\infty$ vs $s_0^2/D_r$.}
\label{fig4}
\end{figure}

We finally show that our results are not specific to soft interaction potentials and confirm 
---this was already shown in Fig.~\ref{fig2}(e)---
that alignment is not necessary to observe extremely deformable active crystals.
We revisited the crystals of active Brownian particles investigated recently by
Paliwal and Dijkstra \cite{paliwal2020role}, which were declared to exhibit a ``defectless hexatic phase", 
a conclusion drawn from positional order correlations decaying faster than a $\eta=\tfrac{1}{3}$ power law.
Particles now interact via only pairwise, hardcore repulsive forces. Angles follow Eq.\eqref{eqtheta} with
$\kappa=0$ and positions are governed by:
\begin{equation}
\dot{\bf r}_i =  s_0 \,{\bf e}(\theta_i) + \mu_r \sum_{j\sim i} \partial_rU(r_{ij}) \,{\bf e}_{ij}
 \label{eqWCA}
\end{equation}
where $U(r)=4\varepsilon[(\tfrac{\sigma}{r})^{12} - (\tfrac{\sigma}{r})^6]+\varepsilon$ is a repulsive Weeks-Chandler-Andersen potential cut off at $r_c=2^{1/6}\sigma$.
Using parameter values similar to those at which ``defectless hexatic order'' was reported but
considering larger systems, we find instead a bona fide crystal phase,
with clear algebraic decay of positional order ---and $\eta$ values easily larger than $\tfrac{1}{3}$---, while 
bond orientational order is truly long-range (Fig.~\ref{fig4})
\footnote{Note that this is only observed beyond a crossover scale 
of the order of 100 lattice steps, a sizeable numerical difficulty, 
which comes in addition to having to use small timesteps.}.
Varying $D_r$ and $s_0$ we find again that $\eta$ depends only on $T_S=s_0^2/2D_r$, 
while $\hat{g}_6^\infty$ does not (insets of Fig.~\ref{fig4}), indicating that, in this case also, $T_6<T_S$.

To summarize, we have shown that fluctuations in 2D active crystals can induce extremely large
deformations without melting: bond order remains long-range, and positional order decays algebraically,
but with an exponent $\eta$ not limited by the $\tfrac{1}{3}$ bound given by KTHNY theory.
For the simple cases studied, linear elastic theory allowed us to rationalize our findings in terms of well-defined 
effective temperatures $T_S$ and $T_6$ 
quantifying respectively large-scale deformations and bond-order fluctuations.
We found in particular that $T_6$, which is expected to control melting, can be much smaller than $T_S$, allowing thus extreme spontaneous deformations while long-range bond-order is preserved.
The root of our results, obtained both with and without weak alignment of the intrinsic axes of active particles,  lies in the sole time-persistence of these polarities~\footnote{The reader might wonder why most of our results
are for the rather 'exotic' case of weak alignment. The main reason is practical: one can use much larger time-steps in this case than with the hardcore WCA potential, saving about one order of magnitude of computation time.}. 
We thus expect similar phenomena to be present in other
types of active crystals, such as those made of chiral or spinning units. 
Generalizing further, the key ingredient can be seen to be the existence of some effective time-correlated local noise.
In line with this idea, preliminary results indicate that even passive crystals submitted to an active bath can display
large spontaneous deformations. 

\acknowledgments
We thank Beno{\^{\i}}t Mahault for a careful reading of the manuscript.
This work is supported by the National Natural Science Foundation of China (Grants No. 12275188 and No. 11922506)

\bibliographystyle{apsrev4-2}
\bibliography{../../Biblio-current.bib}

\begin{thebibliography}{59}%
\makeatletter
\providecommand \@ifxundefined [1]{%
 \@ifx{#1\undefined}
}%
\providecommand \@ifnum [1]{%
 \ifnum #1\expandafter \@firstoftwo
 \else \expandafter \@secondoftwo
 \fi
}%
\providecommand \@ifx [1]{%
 \ifx #1\expandafter \@firstoftwo
 \else \expandafter \@secondoftwo
 \fi
}%
\providecommand \natexlab [1]{#1}%
\providecommand \enquote  [1]{``#1''}%
\providecommand \bibnamefont  [1]{#1}%
\providecommand \bibfnamefont [1]{#1}%
\providecommand \citenamefont [1]{#1}%
\providecommand \href@noop [0]{\@secondoftwo}%
\providecommand \href [0]{\begingroup \@sanitize@url \@href}%
\providecommand \@href[1]{\@@startlink{#1}\@@href}%
\providecommand \@@href[1]{\endgroup#1\@@endlink}%
\providecommand \@sanitize@url [0]{\catcode `\\12\catcode `\$12\catcode
  `\&12\catcode `\#12\catcode `\^12\catcode `\_12\catcode `\%12\relax}%
\providecommand \@@startlink[1]{}%
\providecommand \@@endlink[0]{}%
\providecommand \url  [0]{\begingroup\@sanitize@url \@url }%
\providecommand \@url [1]{\endgroup\@href {#1}{\urlprefix }}%
\providecommand \urlprefix  [0]{URL }%
\providecommand \Eprint [0]{\href }%
\providecommand \doibase [0]{https://doi.org/}%
\providecommand \selectlanguage [0]{\@gobble}%
\providecommand \bibinfo  [0]{\@secondoftwo}%
\providecommand \bibfield  [0]{\@secondoftwo}%
\providecommand \translation [1]{[#1]}%
\providecommand \BibitemOpen [0]{}%
\providecommand \bibitemStop [0]{}%
\providecommand \bibitemNoStop [0]{.\EOS\space}%
\providecommand \EOS [0]{\spacefactor3000\relax}%
\providecommand \BibitemShut  [1]{\csname bibitem#1\endcsname}%
\let\auto@bib@innerbib\@empty
\bibitem [{\citenamefont {{Kosterlitz}}\ and\ \citenamefont
  {{Thouless}}(1972)}]{kosterlitz1972long}%
  \BibitemOpen
  \bibfield  {author} {\bibinfo {author} {\bibfnamefont {J.~M.}\ \bibnamefont
  {{Kosterlitz}}}\ and\ \bibinfo {author} {\bibfnamefont {D.~J.}\ \bibnamefont
  {{Thouless}}},\ }\href {https://doi.org/10.1088/0022-3719/5/11/002}
  {\bibfield  {journal} {\bibinfo  {journal} {Journal of Physics C Solid State
  Physics}\ }\textbf {\bibinfo {volume} {5}},\ \bibinfo {pages} {L124}
  (\bibinfo {year} {1972})}\BibitemShut {NoStop}%
\bibitem [{\citenamefont {Kosterlitz}\ and\ \citenamefont
  {Thouless}(1973)}]{kosterlitz1973ordering}%
  \BibitemOpen
  \bibfield  {author} {\bibinfo {author} {\bibfnamefont {J.~M.}\ \bibnamefont
  {Kosterlitz}}\ and\ \bibinfo {author} {\bibfnamefont {D.~J.}\ \bibnamefont
  {Thouless}},\ }\href
  {http://iopscience.iop.org/article/10.1088/0022-3719/6/7/010/meta} {\bibfield
   {journal} {\bibinfo  {journal} {Journal of Physics C: Solid State Physics}\
  }\textbf {\bibinfo {volume} {6}},\ \bibinfo {pages} {1181} (\bibinfo {year}
  {1973})}\BibitemShut {NoStop}%
\bibitem [{\citenamefont {{Halperin}}\ and\ \citenamefont
  {{Nelson}}(1978)}]{halperin1978theory}%
  \BibitemOpen
  \bibfield  {author} {\bibinfo {author} {\bibfnamefont {B.~I.}\ \bibnamefont
  {{Halperin}}}\ and\ \bibinfo {author} {\bibfnamefont {D.~R.}\ \bibnamefont
  {{Nelson}}},\ }\href {https://doi.org/10.1103/PhysRevLett.41.121} {\bibfield
  {journal} {\bibinfo  {journal} {\prl}\ }\textbf {\bibinfo {volume} {41}},\
  \bibinfo {pages} {121} (\bibinfo {year} {1978})}\BibitemShut {NoStop}%
\bibitem [{\citenamefont {{Nelson}}\ and\ \citenamefont
  {{Halperin}}(1979)}]{nelson1979dislocation}%
  \BibitemOpen
  \bibfield  {author} {\bibinfo {author} {\bibfnamefont {D.~R.}\ \bibnamefont
  {{Nelson}}}\ and\ \bibinfo {author} {\bibfnamefont {B.~I.}\ \bibnamefont
  {{Halperin}}},\ }\href {https://doi.org/10.1103/PhysRevB.19.2457} {\bibfield
  {journal} {\bibinfo  {journal} {\prb}\ }\textbf {\bibinfo {volume} {19}},\
  \bibinfo {pages} {2457} (\bibinfo {year} {1979})}\BibitemShut {NoStop}%
\bibitem [{\citenamefont {{Young}}(1979)}]{young1979melting}%
  \BibitemOpen
  \bibfield  {author} {\bibinfo {author} {\bibfnamefont {A.~P.}\ \bibnamefont
  {{Young}}},\ }\href {https://doi.org/10.1103/PhysRevB.19.1855} {\bibfield
  {journal} {\bibinfo  {journal} {\prb}\ }\textbf {\bibinfo {volume} {19}},\
  \bibinfo {pages} {1855} (\bibinfo {year} {1979})}\BibitemShut {NoStop}%
\bibitem [{\citenamefont {Strandburg}(1988)}]{strandburg1988two-dimensional}%
  \BibitemOpen
  \bibfield  {author} {\bibinfo {author} {\bibfnamefont {K.~J.}\ \bibnamefont
  {Strandburg}},\ }\href {https://doi.org/10.1103/RevModPhys.60.161} {\bibfield
   {journal} {\bibinfo  {journal} {Rev. Mod. Phys.}\ }\textbf {\bibinfo
  {volume} {60}},\ \bibinfo {pages} {161} (\bibinfo {year} {1988})}\BibitemShut
  {NoStop}%
\bibitem [{\citenamefont {{Bialk{\'e}}}\ \emph {et~al.}(2012)\citenamefont
  {{Bialk{\'e}}}, \citenamefont {{Speck}},\ and\ \citenamefont
  {{L{\"o}wen}}}]{bialke2012crystallization}%
  \BibitemOpen
  \bibfield  {author} {\bibinfo {author} {\bibfnamefont {J.}~\bibnamefont
  {{Bialk{\'e}}}}, \bibinfo {author} {\bibfnamefont {T.}~\bibnamefont
  {{Speck}}},\ and\ \bibinfo {author} {\bibfnamefont {H.}~\bibnamefont
  {{L{\"o}wen}}},\ }\href {https://doi.org/10.1103/PhysRevLett.108.168301}
  {\bibfield  {journal} {\bibinfo  {journal} {\prl}\ }\textbf {\bibinfo
  {volume} {108}},\ \bibinfo {eid} {168301} (\bibinfo {year} {2012})},\ \Eprint
  {https://arxiv.org/abs/1112.5281} {arXiv:1112.5281 [cond-mat.soft]}
  \BibitemShut {NoStop}%
\bibitem [{\citenamefont {Redner}\ \emph {et~al.}(2013)\citenamefont {Redner},
  \citenamefont {Hagan},\ and\ \citenamefont {Baskaran}}]{redner2013structure}%
  \BibitemOpen
  \bibfield  {author} {\bibinfo {author} {\bibfnamefont {G.~S.}\ \bibnamefont
  {Redner}}, \bibinfo {author} {\bibfnamefont {M.~F.}\ \bibnamefont {Hagan}},\
  and\ \bibinfo {author} {\bibfnamefont {A.}~\bibnamefont {Baskaran}},\ }\href
  {https://journals.aps.org/prl/abstract/10.1103/PhysRevLett.110.055701}
  {\bibfield  {journal} {\bibinfo  {journal} {Physical Review Letters}\
  }\textbf {\bibinfo {volume} {110}},\ \bibinfo {pages} {055701} (\bibinfo
  {year} {2013})}\BibitemShut {NoStop}%
\bibitem [{\citenamefont {Singh}\ and\ \citenamefont
  {Adhikari}(2016)}]{singh2016universal}%
  \BibitemOpen
  \bibfield  {author} {\bibinfo {author} {\bibfnamefont {R.}~\bibnamefont
  {Singh}}\ and\ \bibinfo {author} {\bibfnamefont {R.}~\bibnamefont
  {Adhikari}},\ }\href {https://doi.org/10.1103/PhysRevLett.117.228002}
  {\bibfield  {journal} {\bibinfo  {journal} {Phys. Rev. Lett.}\ }\textbf
  {\bibinfo {volume} {117}},\ \bibinfo {pages} {228002} (\bibinfo {year}
  {2016})}\BibitemShut {NoStop}%
\bibitem [{\citenamefont {{Thutupalli}}\ \emph {et~al.}(2018)\citenamefont
  {{Thutupalli}}, \citenamefont {{Geyer}}, \citenamefont {{Singh}},
  \citenamefont {{Adhikari}},\ and\ \citenamefont
  {{Stone}}}]{thutupalli2018flow-induced}%
  \BibitemOpen
  \bibfield  {author} {\bibinfo {author} {\bibfnamefont {S.}~\bibnamefont
  {{Thutupalli}}}, \bibinfo {author} {\bibfnamefont {D.}~\bibnamefont
  {{Geyer}}}, \bibinfo {author} {\bibfnamefont {R.}~\bibnamefont {{Singh}}},
  \bibinfo {author} {\bibfnamefont {R.}~\bibnamefont {{Adhikari}}},\ and\
  \bibinfo {author} {\bibfnamefont {H.~A.}\ \bibnamefont {{Stone}}},\ }\href
  {https://doi.org/10.1073/pnas.1718807115} {\bibfield  {journal} {\bibinfo
  {journal} {Proceedings of the National Academy of Science}\ }\textbf
  {\bibinfo {volume} {115}},\ \bibinfo {pages} {5403} (\bibinfo {year}
  {2018})},\ \Eprint {https://arxiv.org/abs/1710.10300} {arXiv:1710.10300
  [cond-mat.soft]} \BibitemShut {NoStop}%
\bibitem [{\citenamefont {Cugliandolo}\ \emph {et~al.}(2017)\citenamefont
  {Cugliandolo}, \citenamefont {Digregorio}, \citenamefont {Gonnella},\ and\
  \citenamefont {Suma}}]{cugliandolo2017phase}%
  \BibitemOpen
  \bibfield  {author} {\bibinfo {author} {\bibfnamefont {L.~F.}\ \bibnamefont
  {Cugliandolo}}, \bibinfo {author} {\bibfnamefont {P.}~\bibnamefont
  {Digregorio}}, \bibinfo {author} {\bibfnamefont {G.}~\bibnamefont
  {Gonnella}},\ and\ \bibinfo {author} {\bibfnamefont {A.}~\bibnamefont
  {Suma}},\ }\href {https://doi.org/10.1103/PhysRevLett.119.268002} {\bibfield
  {journal} {\bibinfo  {journal} {Phys. Rev. Lett.}\ }\textbf {\bibinfo
  {volume} {119}},\ \bibinfo {pages} {268002} (\bibinfo {year}
  {2017})}\BibitemShut {NoStop}%
\bibitem [{\citenamefont {Digregorio}\ \emph {et~al.}(2018)\citenamefont
  {Digregorio}, \citenamefont {Levis}, \citenamefont {Suma}, \citenamefont
  {Cugliandolo}, \citenamefont {Gonnella},\ and\ \citenamefont
  {Pagonabarraga}}]{digregorio2018full}%
  \BibitemOpen
  \bibfield  {author} {\bibinfo {author} {\bibfnamefont {P.}~\bibnamefont
  {Digregorio}}, \bibinfo {author} {\bibfnamefont {D.}~\bibnamefont {Levis}},
  \bibinfo {author} {\bibfnamefont {A.}~\bibnamefont {Suma}}, \bibinfo {author}
  {\bibfnamefont {L.~F.}\ \bibnamefont {Cugliandolo}}, \bibinfo {author}
  {\bibfnamefont {G.}~\bibnamefont {Gonnella}},\ and\ \bibinfo {author}
  {\bibfnamefont {I.}~\bibnamefont {Pagonabarraga}},\ }\href
  {https://doi.org/10.1103/PhysRevLett.121.098003} {\bibfield  {journal}
  {\bibinfo  {journal} {Phys. Rev. Lett.}\ }\textbf {\bibinfo {volume} {121}},\
  \bibinfo {pages} {098003} (\bibinfo {year} {2018})}\BibitemShut {NoStop}%
\bibitem [{\citenamefont {{Digregorio}}\ \emph {et~al.}(2019)\citenamefont
  {{Digregorio}}, \citenamefont {{Levis}}, \citenamefont {{Suma}},
  \citenamefont {{Cugliandolo}}, \citenamefont {{Gonnella}},\ and\
  \citenamefont {{Pagonabarraga}}}]{digregorio20192D}%
  \BibitemOpen
  \bibfield  {author} {\bibinfo {author} {\bibfnamefont {P.}~\bibnamefont
  {{Digregorio}}}, \bibinfo {author} {\bibfnamefont {D.}~\bibnamefont
  {{Levis}}}, \bibinfo {author} {\bibfnamefont {A.}~\bibnamefont {{Suma}}},
  \bibinfo {author} {\bibfnamefont {L.~F.}\ \bibnamefont {{Cugliandolo}}},
  \bibinfo {author} {\bibfnamefont {G.}~\bibnamefont {{Gonnella}}},\ and\
  \bibinfo {author} {\bibfnamefont {I.}~\bibnamefont {{Pagonabarraga}}},\ }in\
  \href {https://doi.org/10.1088/1742-6596/1163/1/012073} {\emph {\bibinfo
  {booktitle} {Journal of Physics Conference Series}}},\ \bibinfo {series}
  {Journal of Physics Conference Series}, Vol.\ \bibinfo {volume} {1163}\
  (\bibinfo {year} {2019})\ p.\ \bibinfo {pages} {012073}\BibitemShut {NoStop}%
\bibitem [{\citenamefont {{Klamser}}\ \emph {et~al.}(2018)\citenamefont
  {{Klamser}}, \citenamefont {{Kapfer}},\ and\ \citenamefont
  {{Krauth}}}]{klamser2018thermodynamic}%
  \BibitemOpen
  \bibfield  {author} {\bibinfo {author} {\bibfnamefont {J.~U.}\ \bibnamefont
  {{Klamser}}}, \bibinfo {author} {\bibfnamefont {S.~C.}\ \bibnamefont
  {{Kapfer}}},\ and\ \bibinfo {author} {\bibfnamefont {W.}~\bibnamefont
  {{Krauth}}},\ }\href {https://doi.org/10.1038/s41467-018-07491-5} {\bibfield
  {journal} {\bibinfo  {journal} {Nature Communications}\ }\textbf {\bibinfo
  {volume} {9}},\ \bibinfo {eid} {5045} (\bibinfo {year} {2018})},\ \Eprint
  {https://arxiv.org/abs/1802.10021} {arXiv:1802.10021 [cond-mat.stat-mech]}
  \BibitemShut {NoStop}%
\bibitem [{\citenamefont {{Pasupalak}}\ \emph {et~al.}(2020)\citenamefont
  {{Pasupalak}}, \citenamefont {{Yan-Wei}}, \citenamefont {{Ni}},\ and\
  \citenamefont {{Pica Ciamarra}}}]{pasupalak2020hexatic}%
  \BibitemOpen
  \bibfield  {author} {\bibinfo {author} {\bibfnamefont {A.}~\bibnamefont
  {{Pasupalak}}}, \bibinfo {author} {\bibfnamefont {L.}~\bibnamefont
  {{Yan-Wei}}}, \bibinfo {author} {\bibfnamefont {R.}~\bibnamefont {{Ni}}},\
  and\ \bibinfo {author} {\bibfnamefont {M.}~\bibnamefont {{Pica Ciamarra}}},\
  }\href {https://doi.org/10.1039/D0SM00109K} {\bibfield  {journal} {\bibinfo
  {journal} {Soft Matter}\ }\textbf {\bibinfo {volume} {16}},\ \bibinfo {pages}
  {3914} (\bibinfo {year} {2020})}\BibitemShut {NoStop}%
\bibitem [{\citenamefont {Paliwal}\ and\ \citenamefont
  {Dijkstra}(2020)}]{paliwal2020role}%
  \BibitemOpen
  \bibfield  {author} {\bibinfo {author} {\bibfnamefont {S.}~\bibnamefont
  {Paliwal}}\ and\ \bibinfo {author} {\bibfnamefont {M.}~\bibnamefont
  {Dijkstra}},\ }\href {https://doi.org/10.1103/PhysRevResearch.2.012013}
  {\bibfield  {journal} {\bibinfo  {journal} {Phys. Rev. Research}\ }\textbf
  {\bibinfo {volume} {2}},\ \bibinfo {pages} {012013} (\bibinfo {year}
  {2020})}\BibitemShut {NoStop}%
\bibitem [{\citenamefont {Loewe}\ \emph {et~al.}(2020)\citenamefont {Loewe},
  \citenamefont {Chiang}, \citenamefont {Marenduzzo},\ and\ \citenamefont
  {Marchetti}}]{loewe2020solid}%
  \BibitemOpen
  \bibfield  {author} {\bibinfo {author} {\bibfnamefont {B.}~\bibnamefont
  {Loewe}}, \bibinfo {author} {\bibfnamefont {M.}~\bibnamefont {Chiang}},
  \bibinfo {author} {\bibfnamefont {D.}~\bibnamefont {Marenduzzo}},\ and\
  \bibinfo {author} {\bibfnamefont {M.~C.}\ \bibnamefont {Marchetti}},\ }\href
  {https://doi.org/10.1103/PhysRevLett.125.038003} {\bibfield  {journal}
  {\bibinfo  {journal} {Phys. Rev. Lett.}\ }\textbf {\bibinfo {volume} {125}},\
  \bibinfo {pages} {038003} (\bibinfo {year} {2020})}\BibitemShut {NoStop}%
\bibitem [{\citenamefont {Digregorio}\ \emph {et~al.}(2022)\citenamefont
  {Digregorio}, \citenamefont {Levis}, \citenamefont {Cugliandolo},
  \citenamefont {Gonnella},\ and\ \citenamefont
  {Pagonabarraga}}]{digregorio2022unified}%
  \BibitemOpen
  \bibfield  {author} {\bibinfo {author} {\bibfnamefont {P.}~\bibnamefont
  {Digregorio}}, \bibinfo {author} {\bibfnamefont {D.}~\bibnamefont {Levis}},
  \bibinfo {author} {\bibfnamefont {L.~F.}\ \bibnamefont {Cugliandolo}},
  \bibinfo {author} {\bibfnamefont {G.}~\bibnamefont {Gonnella}},\ and\
  \bibinfo {author} {\bibfnamefont {I.}~\bibnamefont {Pagonabarraga}},\ }\href
  {https://doi.org/10.1039/D1SM01411K} {\bibfield  {journal} {\bibinfo
  {journal} {Soft Matter}\ }\textbf {\bibinfo {volume} {18}},\ \bibinfo {pages}
  {566} (\bibinfo {year} {2022})}\BibitemShut {NoStop}%
\bibitem [{\citenamefont {{Riedel}}\ \emph {et~al.}(2005)\citenamefont
  {{Riedel}}, \citenamefont {{Kruse}},\ and\ \citenamefont
  {{Howard}}}]{riedel2005self-organized}%
  \BibitemOpen
  \bibfield  {author} {\bibinfo {author} {\bibfnamefont {I.~H.}\ \bibnamefont
  {{Riedel}}}, \bibinfo {author} {\bibfnamefont {K.}~\bibnamefont {{Kruse}}},\
  and\ \bibinfo {author} {\bibfnamefont {J.}~\bibnamefont {{Howard}}},\ }\href
  {https://doi.org/10.1126/science.1110329} {\bibfield  {journal} {\bibinfo
  {journal} {Science}\ }\textbf {\bibinfo {volume} {309}},\ \bibinfo {pages}
  {300} (\bibinfo {year} {2005})}\BibitemShut {NoStop}%
\bibitem [{\citenamefont {Sumino}\ \emph {et~al.}(2012)\citenamefont {Sumino},
  \citenamefont {Nagai}, \citenamefont {Shitaka}, \citenamefont {Tanaka},
  \citenamefont {Yoshikawa}, \citenamefont {Chat{\'e}},\ and\ \citenamefont
  {Oiwa}}]{sumino2012large}%
  \BibitemOpen
  \bibfield  {author} {\bibinfo {author} {\bibfnamefont {Y.}~\bibnamefont
  {Sumino}}, \bibinfo {author} {\bibfnamefont {K.~H.}\ \bibnamefont {Nagai}},
  \bibinfo {author} {\bibfnamefont {Y.}~\bibnamefont {Shitaka}}, \bibinfo
  {author} {\bibfnamefont {D.}~\bibnamefont {Tanaka}}, \bibinfo {author}
  {\bibfnamefont {K.}~\bibnamefont {Yoshikawa}}, \bibinfo {author}
  {\bibfnamefont {H.}~\bibnamefont {Chat{\'e}}},\ and\ \bibinfo {author}
  {\bibfnamefont {K.}~\bibnamefont {Oiwa}},\ }\href
  {https://www.nature.com/articles/nature10874} {\bibfield  {journal} {\bibinfo
   {journal} {Nature}\ }\textbf {\bibinfo {volume} {483}},\ \bibinfo {pages}
  {448} (\bibinfo {year} {2012})}\BibitemShut {NoStop}%
\bibitem [{\citenamefont {{Nguyen}}\ \emph {et~al.}(2014)\citenamefont
  {{Nguyen}}, \citenamefont {{Klotsa}}, \citenamefont {{Engel}},\ and\
  \citenamefont {{Glotzer}}}]{nguyen2014emergent}%
  \BibitemOpen
  \bibfield  {author} {\bibinfo {author} {\bibfnamefont {N.~H.~P.}\
  \bibnamefont {{Nguyen}}}, \bibinfo {author} {\bibfnamefont {D.}~\bibnamefont
  {{Klotsa}}}, \bibinfo {author} {\bibfnamefont {M.}~\bibnamefont {{Engel}}},\
  and\ \bibinfo {author} {\bibfnamefont {S.~C.}\ \bibnamefont {{Glotzer}}},\
  }\href {https://doi.org/10.1103/PhysRevLett.112.075701} {\bibfield  {journal}
  {\bibinfo  {journal} {\prl}\ }\textbf {\bibinfo {volume} {112}},\ \bibinfo
  {eid} {075701} (\bibinfo {year} {2014})},\ \Eprint
  {https://arxiv.org/abs/1308.2219} {arXiv:1308.2219 [cond-mat.soft]}
  \BibitemShut {NoStop}%
\bibitem [{\citenamefont {{Yeo}}\ \emph {et~al.}(2015)\citenamefont {{Yeo}},
  \citenamefont {{Lushi}},\ and\ \citenamefont
  {{Vlahovska}}}]{yeo2015collective}%
  \BibitemOpen
  \bibfield  {author} {\bibinfo {author} {\bibfnamefont {K.}~\bibnamefont
  {{Yeo}}}, \bibinfo {author} {\bibfnamefont {E.}~\bibnamefont {{Lushi}}},\
  and\ \bibinfo {author} {\bibfnamefont {P.~M.}\ \bibnamefont {{Vlahovska}}},\
  }\href {https://doi.org/10.1103/PhysRevLett.114.188301} {\bibfield  {journal}
  {\bibinfo  {journal} {\prl}\ }\textbf {\bibinfo {volume} {114}},\ \bibinfo
  {eid} {188301} (\bibinfo {year} {2015})},\ \Eprint
  {https://arxiv.org/abs/1410.2878} {arXiv:1410.2878 [cond-mat.soft]}
  \BibitemShut {NoStop}%
\bibitem [{\citenamefont {{Goto}}\ and\ \citenamefont
  {{Tanaka}}(2015)}]{goto2015purely}%
  \BibitemOpen
  \bibfield  {author} {\bibinfo {author} {\bibfnamefont {Y.}~\bibnamefont
  {{Goto}}}\ and\ \bibinfo {author} {\bibfnamefont {H.}~\bibnamefont
  {{Tanaka}}},\ }\href {https://doi.org/10.1038/ncomms6994} {\bibfield
  {journal} {\bibinfo  {journal} {Nature Communications}\ }\textbf {\bibinfo
  {volume} {6}},\ \bibinfo {eid} {5994} (\bibinfo {year} {2015})},\ \Eprint
  {https://arxiv.org/abs/1502.05174} {arXiv:1502.05174 [cond-mat.soft]}
  \BibitemShut {NoStop}%
\bibitem [{\citenamefont {{Petroff}}\ \emph {et~al.}(2015)\citenamefont
  {{Petroff}}, \citenamefont {{Wu}},\ and\ \citenamefont
  {{Libchaber}}}]{petroff2015fast-moving}%
  \BibitemOpen
  \bibfield  {author} {\bibinfo {author} {\bibfnamefont {A.~P.}\ \bibnamefont
  {{Petroff}}}, \bibinfo {author} {\bibfnamefont {X.-L.}\ \bibnamefont
  {{Wu}}},\ and\ \bibinfo {author} {\bibfnamefont {A.}~\bibnamefont
  {{Libchaber}}},\ }\href {https://doi.org/10.1103/PhysRevLett.114.158102}
  {\bibfield  {journal} {\bibinfo  {journal} {\prl}\ }\textbf {\bibinfo
  {volume} {114}},\ \bibinfo {eid} {158102} (\bibinfo {year}
  {2015})}\BibitemShut {NoStop}%
\bibitem [{\citenamefont {Oppenheimer}\ \emph {et~al.}(2019)\citenamefont
  {Oppenheimer}, \citenamefont {Stein},\ and\ \citenamefont
  {Shelley}}]{oppenheimer2019rotating}%
  \BibitemOpen
  \bibfield  {author} {\bibinfo {author} {\bibfnamefont {N.}~\bibnamefont
  {Oppenheimer}}, \bibinfo {author} {\bibfnamefont {D.~B.}\ \bibnamefont
  {Stein}},\ and\ \bibinfo {author} {\bibfnamefont {M.~J.}\ \bibnamefont
  {Shelley}},\ }\href {https://doi.org/10.1103/PhysRevLett.123.148101}
  {\bibfield  {journal} {\bibinfo  {journal} {Phys. Rev. Lett.}\ }\textbf
  {\bibinfo {volume} {123}},\ \bibinfo {pages} {148101} (\bibinfo {year}
  {2019})}\BibitemShut {NoStop}%
\bibitem [{\citenamefont {{Huang}}\ \emph {et~al.}(2020)\citenamefont
  {{Huang}}, \citenamefont {{Menzel}},\ and\ \citenamefont
  {{L{\"o}wen}}}]{huang2020dynamical}%
  \BibitemOpen
  \bibfield  {author} {\bibinfo {author} {\bibfnamefont {Z.-F.}\ \bibnamefont
  {{Huang}}}, \bibinfo {author} {\bibfnamefont {A.~M.}\ \bibnamefont
  {{Menzel}}},\ and\ \bibinfo {author} {\bibfnamefont {H.}~\bibnamefont
  {{L{\"o}wen}}},\ }\href {https://doi.org/10.1103/PhysRevLett.125.218002}
  {\bibfield  {journal} {\bibinfo  {journal} {\prl}\ }\textbf {\bibinfo
  {volume} {125}},\ \bibinfo {eid} {218002} (\bibinfo {year} {2020})},\ \Eprint
  {https://arxiv.org/abs/2010.01252} {arXiv:2010.01252 [cond-mat.soft]}
  \BibitemShut {NoStop}%
\bibitem [{\citenamefont {{James}}\ \emph {et~al.}(2021)\citenamefont
  {{James}}, \citenamefont {{Suchla}}, \citenamefont {{Dunkel}},\ and\
  \citenamefont {{Wilczek}}}]{james2021emergence}%
  \BibitemOpen
  \bibfield  {author} {\bibinfo {author} {\bibfnamefont {M.}~\bibnamefont
  {{James}}}, \bibinfo {author} {\bibfnamefont {D.~A.}\ \bibnamefont
  {{Suchla}}}, \bibinfo {author} {\bibfnamefont {J.}~\bibnamefont {{Dunkel}}},\
  and\ \bibinfo {author} {\bibfnamefont {M.}~\bibnamefont {{Wilczek}}},\ }\href
  {https://doi.org/10.1038/s41467-021-25545-z} {\bibfield  {journal} {\bibinfo
  {journal} {Nature Communications}\ }\textbf {\bibinfo {volume} {12}},\
  \bibinfo {eid} {5630} (\bibinfo {year} {2021})},\ \Eprint
  {https://arxiv.org/abs/2005.06217} {arXiv:2005.06217 [cond-mat.soft]}
  \BibitemShut {NoStop}%
\bibitem [{\citenamefont {{Tan}}\ \emph {et~al.}(2022)\citenamefont {{Tan}},
  \citenamefont {{Mietke}}, \citenamefont {{Li}}, \citenamefont {{Chen}},
  \citenamefont {{Higinbotham}}, \citenamefont {{Foster}}, \citenamefont
  {{Gokhale}}, \citenamefont {{Dunkel}},\ and\ \citenamefont
  {{Fakhri}}}]{tan2022odd}%
  \BibitemOpen
  \bibfield  {author} {\bibinfo {author} {\bibfnamefont {T.~H.}\ \bibnamefont
  {{Tan}}}, \bibinfo {author} {\bibfnamefont {A.}~\bibnamefont {{Mietke}}},
  \bibinfo {author} {\bibfnamefont {J.}~\bibnamefont {{Li}}}, \bibinfo {author}
  {\bibfnamefont {Y.}~\bibnamefont {{Chen}}}, \bibinfo {author} {\bibfnamefont
  {H.}~\bibnamefont {{Higinbotham}}}, \bibinfo {author} {\bibfnamefont {P.~J.}\
  \bibnamefont {{Foster}}}, \bibinfo {author} {\bibfnamefont {S.}~\bibnamefont
  {{Gokhale}}}, \bibinfo {author} {\bibfnamefont {J.}~\bibnamefont
  {{Dunkel}}},\ and\ \bibinfo {author} {\bibfnamefont {N.}~\bibnamefont
  {{Fakhri}}},\ }\href {https://doi.org/10.1038/s41586-022-04889-6} {\bibfield
  {journal} {\bibinfo  {journal} {\nat}\ }\textbf {\bibinfo {volume} {607}},\
  \bibinfo {pages} {287} (\bibinfo {year} {2022})},\ \Eprint
  {https://arxiv.org/abs/2105.07507} {arXiv:2105.07507 [cond-mat.soft]}
  \BibitemShut {NoStop}%
\bibitem [{\citenamefont {Oppenheimer}\ \emph {et~al.}(2022)\citenamefont
  {Oppenheimer}, \citenamefont {Stein}, \citenamefont {Zion},\ and\
  \citenamefont {Shelley}}]{oppenheimer2022hyperuniformity}%
  \BibitemOpen
  \bibfield  {author} {\bibinfo {author} {\bibfnamefont {N.}~\bibnamefont
  {Oppenheimer}}, \bibinfo {author} {\bibfnamefont {D.~B.}\ \bibnamefont
  {Stein}}, \bibinfo {author} {\bibfnamefont {M.~Y.~B.}\ \bibnamefont {Zion}},\
  and\ \bibinfo {author} {\bibfnamefont {M.~J.}\ \bibnamefont {Shelley}},\
  }\href {https://doi.org/10.1038/s41467-022-28375-9} {\bibfield  {journal}
  {\bibinfo  {journal} {Nature Communications}\ }\textbf {\bibinfo {volume}
  {13}},\ \bibinfo {pages} {804} (\bibinfo {year} {2022})}\BibitemShut
  {NoStop}%
\bibitem [{\citenamefont {{van Zuiden}}\ \emph {et~al.}(2016)\citenamefont
  {{van Zuiden}}, \citenamefont {{Paulose}}, \citenamefont {{Irvine}},
  \citenamefont {{Bartolo}},\ and\ \citenamefont
  {{Vitelli}}}]{vanzuiden2016spatiotemporal}%
  \BibitemOpen
  \bibfield  {author} {\bibinfo {author} {\bibfnamefont {B.~C.}\ \bibnamefont
  {{van Zuiden}}}, \bibinfo {author} {\bibfnamefont {J.}~\bibnamefont
  {{Paulose}}}, \bibinfo {author} {\bibfnamefont {W.~T.~M.}\ \bibnamefont
  {{Irvine}}}, \bibinfo {author} {\bibfnamefont {D.}~\bibnamefont
  {{Bartolo}}},\ and\ \bibinfo {author} {\bibfnamefont {V.}~\bibnamefont
  {{Vitelli}}},\ }\href {https://doi.org/10.1073/pnas.1609572113} {\bibfield
  {journal} {\bibinfo  {journal} {Proceedings of the National Academy of
  Science}\ }\textbf {\bibinfo {volume} {113}},\ \bibinfo {pages} {12919}
  (\bibinfo {year} {2016})},\ \Eprint {https://arxiv.org/abs/1606.03934}
  {arXiv:1606.03934 [cond-mat.soft]} \BibitemShut {NoStop}%
\bibitem [{\citenamefont {{Scheibner}}\ \emph {et~al.}(2020)\citenamefont
  {{Scheibner}}, \citenamefont {{Souslov}}, \citenamefont {{Banerjee}},
  \citenamefont {{Sur{\'o}wka}}, \citenamefont {{Irvine}},\ and\ \citenamefont
  {{Vitelli}}}]{scheibner2020odd}%
  \BibitemOpen
  \bibfield  {author} {\bibinfo {author} {\bibfnamefont {C.}~\bibnamefont
  {{Scheibner}}}, \bibinfo {author} {\bibfnamefont {A.}~\bibnamefont
  {{Souslov}}}, \bibinfo {author} {\bibfnamefont {D.}~\bibnamefont
  {{Banerjee}}}, \bibinfo {author} {\bibfnamefont {P.}~\bibnamefont
  {{Sur{\'o}wka}}}, \bibinfo {author} {\bibfnamefont {W.~T.~M.}\ \bibnamefont
  {{Irvine}}},\ and\ \bibinfo {author} {\bibfnamefont {V.}~\bibnamefont
  {{Vitelli}}},\ }\href {https://doi.org/10.1038/s41567-020-0795-y} {\bibfield
  {journal} {\bibinfo  {journal} {Nature Physics}\ }\textbf {\bibinfo {volume}
  {16}},\ \bibinfo {pages} {475} (\bibinfo {year} {2020})},\ \Eprint
  {https://arxiv.org/abs/1902.07760} {arXiv:1902.07760 [cond-mat.soft]}
  \BibitemShut {NoStop}%
\bibitem [{\citenamefont {Bililign}\ \emph {et~al.}(2022)\citenamefont
  {Bililign}, \citenamefont {Balboa~Usabiaga}, \citenamefont {Ganan},
  \citenamefont {Poncet}, \citenamefont {Soni}, \citenamefont {Magkiriadou},
  \citenamefont {Shelley}, \citenamefont {Bartolo},\ and\ \citenamefont
  {Irvine}}]{bililign2022motile}%
  \BibitemOpen
  \bibfield  {author} {\bibinfo {author} {\bibfnamefont {E.~S.}\ \bibnamefont
  {Bililign}}, \bibinfo {author} {\bibfnamefont {F.}~\bibnamefont
  {Balboa~Usabiaga}}, \bibinfo {author} {\bibfnamefont {Y.~A.}\ \bibnamefont
  {Ganan}}, \bibinfo {author} {\bibfnamefont {A.}~\bibnamefont {Poncet}},
  \bibinfo {author} {\bibfnamefont {V.}~\bibnamefont {Soni}}, \bibinfo {author}
  {\bibfnamefont {S.}~\bibnamefont {Magkiriadou}}, \bibinfo {author}
  {\bibfnamefont {M.~J.}\ \bibnamefont {Shelley}}, \bibinfo {author}
  {\bibfnamefont {D.}~\bibnamefont {Bartolo}},\ and\ \bibinfo {author}
  {\bibfnamefont {W.~T.~M.}\ \bibnamefont {Irvine}},\ }\href
  {https://doi.org/10.1038/s41567-021-01429-3} {\bibfield  {journal} {\bibinfo
  {journal} {Nature Physics}\ }\textbf {\bibinfo {volume} {18}},\ \bibinfo
  {pages} {212} (\bibinfo {year} {2022})}\BibitemShut {NoStop}%
\bibitem [{\citenamefont {Braverman}\ \emph {et~al.}(2021)\citenamefont
  {Braverman}, \citenamefont {Scheibner}, \citenamefont {VanSaders},\ and\
  \citenamefont {Vitelli}}]{braverman2021topological}%
  \BibitemOpen
  \bibfield  {author} {\bibinfo {author} {\bibfnamefont {L.}~\bibnamefont
  {Braverman}}, \bibinfo {author} {\bibfnamefont {C.}~\bibnamefont
  {Scheibner}}, \bibinfo {author} {\bibfnamefont {B.}~\bibnamefont
  {VanSaders}},\ and\ \bibinfo {author} {\bibfnamefont {V.}~\bibnamefont
  {Vitelli}},\ }\href {https://doi.org/10.1103/PhysRevLett.127.268001}
  {\bibfield  {journal} {\bibinfo  {journal} {Phys. Rev. Lett.}\ }\textbf
  {\bibinfo {volume} {127}},\ \bibinfo {pages} {268001} (\bibinfo {year}
  {2021})}\BibitemShut {NoStop}%
\bibitem [{\citenamefont {{Gr{\'e}goire}}\ \emph {et~al.}(2003)\citenamefont
  {{Gr{\'e}goire}}, \citenamefont {{Chat{\'e}}},\ and\ \citenamefont
  {{Tu}}}]{gregoire2003moving}%
  \BibitemOpen
  \bibfield  {author} {\bibinfo {author} {\bibfnamefont {G.}~\bibnamefont
  {{Gr{\'e}goire}}}, \bibinfo {author} {\bibfnamefont {H.}~\bibnamefont
  {{Chat{\'e}}}},\ and\ \bibinfo {author} {\bibfnamefont {Y.}~\bibnamefont
  {{Tu}}},\ }\href {https://doi.org/10.1016/S0167-2789(03)00102-7} {\bibfield
  {journal} {\bibinfo  {journal} {Physica D Nonlinear Phenomena}\ }\textbf
  {\bibinfo {volume} {181}},\ \bibinfo {pages} {157} (\bibinfo {year}
  {2003})},\ \Eprint {https://arxiv.org/abs/cond-mat/0401257}
  {arXiv:cond-mat/0401257 [cond-mat.stat-mech]} \BibitemShut {NoStop}%
\bibitem [{\citenamefont {Ferrante}\ \emph {et~al.}(2013)\citenamefont
  {Ferrante}, \citenamefont {Turgut}, \citenamefont {Dorigo},\ and\
  \citenamefont {Huepe}}]{ferrante2013elasticity}%
  \BibitemOpen
  \bibfield  {author} {\bibinfo {author} {\bibfnamefont {E.}~\bibnamefont
  {Ferrante}}, \bibinfo {author} {\bibfnamefont {A.~E.}\ \bibnamefont
  {Turgut}}, \bibinfo {author} {\bibfnamefont {M.}~\bibnamefont {Dorigo}},\
  and\ \bibinfo {author} {\bibfnamefont {C.}~\bibnamefont {Huepe}},\ }\href
  {https://doi.org/10.1103/PhysRevLett.111.268302} {\bibfield  {journal}
  {\bibinfo  {journal} {Phys. Rev. Lett.}\ }\textbf {\bibinfo {volume} {111}},\
  \bibinfo {pages} {268302} (\bibinfo {year} {2013})}\BibitemShut {NoStop}%
\bibitem [{\citenamefont {{Ferrante}}\ \emph {et~al.}(2013)\citenamefont
  {{Ferrante}}, \citenamefont {{Emre Turgut}}, \citenamefont {{Dorigo}},\ and\
  \citenamefont {{Huepe}}}]{ferrante2013collective}%
  \BibitemOpen
  \bibfield  {author} {\bibinfo {author} {\bibfnamefont {E.}~\bibnamefont
  {{Ferrante}}}, \bibinfo {author} {\bibfnamefont {A.}~\bibnamefont {{Emre
  Turgut}}}, \bibinfo {author} {\bibfnamefont {M.}~\bibnamefont {{Dorigo}}},\
  and\ \bibinfo {author} {\bibfnamefont {C.}~\bibnamefont {{Huepe}}},\ }\href
  {https://doi.org/10.1088/1367-2630/15/9/095011} {\bibfield  {journal}
  {\bibinfo  {journal} {New Journal of Physics}\ }\textbf {\bibinfo {volume}
  {15}},\ \bibinfo {eid} {095011} (\bibinfo {year} {2013})}\BibitemShut
  {NoStop}%
\bibitem [{\citenamefont {{Menzel}}(2013)}]{menzel2013unidirectional}%
  \BibitemOpen
  \bibfield  {author} {\bibinfo {author} {\bibfnamefont {A.~M.}\ \bibnamefont
  {{Menzel}}},\ }\href {https://doi.org/10.1088/0953-8984/25/50/505103}
  {\bibfield  {journal} {\bibinfo  {journal} {Journal of Physics Condensed
  Matter}\ }\textbf {\bibinfo {volume} {25}},\ \bibinfo {eid} {505103}
  (\bibinfo {year} {2013})},\ \Eprint {https://arxiv.org/abs/1310.7806}
  {arXiv:1310.7806 [cond-mat.soft]} \BibitemShut {NoStop}%
\bibitem [{\citenamefont {{Menzel}}\ and\ \citenamefont
  {{L{\"o}wen}}(2013)}]{menzel2013traveling}%
  \BibitemOpen
  \bibfield  {author} {\bibinfo {author} {\bibfnamefont {A.~M.}\ \bibnamefont
  {{Menzel}}}\ and\ \bibinfo {author} {\bibfnamefont {H.}~\bibnamefont
  {{L{\"o}wen}}},\ }\href {https://doi.org/10.1103/PhysRevLett.110.055702}
  {\bibfield  {journal} {\bibinfo  {journal} {\prl}\ }\textbf {\bibinfo
  {volume} {110}},\ \bibinfo {eid} {055702} (\bibinfo {year} {2013})},\ \Eprint
  {https://arxiv.org/abs/1209.3537} {arXiv:1209.3537 [cond-mat.soft]}
  \BibitemShut {NoStop}%
\bibitem [{\citenamefont {{Menzel}}\ \emph {et~al.}(2014)\citenamefont
  {{Menzel}}, \citenamefont {{Ohta}},\ and\ \citenamefont
  {{L{\"o}wen}}}]{menzel2014active}%
  \BibitemOpen
  \bibfield  {author} {\bibinfo {author} {\bibfnamefont {A.~M.}\ \bibnamefont
  {{Menzel}}}, \bibinfo {author} {\bibfnamefont {T.}~\bibnamefont {{Ohta}}},\
  and\ \bibinfo {author} {\bibfnamefont {H.}~\bibnamefont {{L{\"o}wen}}},\
  }\href {https://doi.org/10.1103/PhysRevE.89.022301} {\bibfield  {journal}
  {\bibinfo  {journal} {\pre}\ }\textbf {\bibinfo {volume} {89}},\ \bibinfo
  {eid} {022301} (\bibinfo {year} {2014})},\ \Eprint
  {https://arxiv.org/abs/1401.5332} {arXiv:1401.5332 [cond-mat.soft]}
  \BibitemShut {NoStop}%
\bibitem [{\citenamefont {{Ophaus}}\ \emph {et~al.}(2018)\citenamefont
  {{Ophaus}}, \citenamefont {{Gurevich}},\ and\ \citenamefont
  {{Thiele}}}]{ophaus2018resting}%
  \BibitemOpen
  \bibfield  {author} {\bibinfo {author} {\bibfnamefont {L.}~\bibnamefont
  {{Ophaus}}}, \bibinfo {author} {\bibfnamefont {S.~V.}\ \bibnamefont
  {{Gurevich}}},\ and\ \bibinfo {author} {\bibfnamefont {U.}~\bibnamefont
  {{Thiele}}},\ }\href {https://doi.org/10.1103/PhysRevE.98.022608} {\bibfield
  {journal} {\bibinfo  {journal} {\pre}\ }\textbf {\bibinfo {volume} {98}},\
  \bibinfo {eid} {022608} (\bibinfo {year} {2018})},\ \Eprint
  {https://arxiv.org/abs/1803.08902} {arXiv:1803.08902 [nlin.PS]} \BibitemShut
  {NoStop}%
\bibitem [{\citenamefont {{Alaimo}}\ \emph {et~al.}(2016)\citenamefont
  {{Alaimo}}, \citenamefont {{Praetorius}},\ and\ \citenamefont
  {{Voigt}}}]{alaimo2016microscopic}%
  \BibitemOpen
  \bibfield  {author} {\bibinfo {author} {\bibfnamefont {F.}~\bibnamefont
  {{Alaimo}}}, \bibinfo {author} {\bibfnamefont {S.}~\bibnamefont
  {{Praetorius}}},\ and\ \bibinfo {author} {\bibfnamefont {A.}~\bibnamefont
  {{Voigt}}},\ }\href {https://doi.org/10.1088/1367-2630/18/8/083008}
  {\bibfield  {journal} {\bibinfo  {journal} {New Journal of Physics}\ }\textbf
  {\bibinfo {volume} {18}},\ \bibinfo {eid} {083008} (\bibinfo {year}
  {2016})},\ \Eprint {https://arxiv.org/abs/1604.06694} {arXiv:1604.06694
  [cond-mat.soft]} \BibitemShut {NoStop}%
\bibitem [{\citenamefont {{Weber}}\ \emph {et~al.}(2014)\citenamefont
  {{Weber}}, \citenamefont {{Bock}},\ and\ \citenamefont
  {{Frey}}}]{weber2014defect-mediated}%
  \BibitemOpen
  \bibfield  {author} {\bibinfo {author} {\bibfnamefont {C.~A.}\ \bibnamefont
  {{Weber}}}, \bibinfo {author} {\bibfnamefont {C.}~\bibnamefont {{Bock}}},\
  and\ \bibinfo {author} {\bibfnamefont {E.}~\bibnamefont {{Frey}}},\ }\href
  {https://doi.org/10.1103/PhysRevLett.112.168301} {\bibfield  {journal}
  {\bibinfo  {journal} {\prl}\ }\textbf {\bibinfo {volume} {112}},\ \bibinfo
  {eid} {168301} (\bibinfo {year} {2014})},\ \Eprint
  {https://arxiv.org/abs/1404.1619} {arXiv:1404.1619 [cond-mat.soft]}
  \BibitemShut {NoStop}%
\bibitem [{\citenamefont {{Rana}}\ \emph {et~al.}(2019)\citenamefont {{Rana}},
  \citenamefont {{Samsuzzaman}},\ and\ \citenamefont
  {{Saha}}}]{rana2019tuning}%
  \BibitemOpen
  \bibfield  {author} {\bibinfo {author} {\bibfnamefont {S.}~\bibnamefont
  {{Rana}}}, \bibinfo {author} {\bibfnamefont {M.}~\bibnamefont
  {{Samsuzzaman}}},\ and\ \bibinfo {author} {\bibfnamefont {A.}~\bibnamefont
  {{Saha}}},\ }\href {https://doi.org/10.1039/C9SM01691K} {\bibfield  {journal}
  {\bibinfo  {journal} {Soft Matter}\ }\textbf {\bibinfo {volume} {15}},\
  \bibinfo {pages} {8865} (\bibinfo {year} {2019})},\ \Eprint
  {https://arxiv.org/abs/1903.12241} {arXiv:1903.12241 [cond-mat.soft]}
  \BibitemShut {NoStop}%
\bibitem [{\citenamefont {{Huang}}\ \emph {et~al.}(2021)\citenamefont
  {{Huang}}, \citenamefont {{Chen}},\ and\ \citenamefont
  {{Xing}}}]{huang2021alignment}%
  \BibitemOpen
  \bibfield  {author} {\bibinfo {author} {\bibfnamefont {C.}~\bibnamefont
  {{Huang}}}, \bibinfo {author} {\bibfnamefont {L.}~\bibnamefont {{Chen}}},\
  and\ \bibinfo {author} {\bibfnamefont {X.}~\bibnamefont {{Xing}}},\ }\href
  {https://doi.org/10.1103/PhysRevE.104.064605} {\bibfield  {journal} {\bibinfo
   {journal} {\pre}\ }\textbf {\bibinfo {volume} {104}},\ \bibinfo {eid}
  {064605} (\bibinfo {year} {2021})},\ \Eprint
  {https://arxiv.org/abs/2105.14535} {arXiv:2105.14535 [cond-mat.soft]}
  \BibitemShut {NoStop}%
\bibitem [{\citenamefont {{Briand}}\ \emph {et~al.}(2018)\citenamefont
  {{Briand}}, \citenamefont {{Schindler}},\ and\ \citenamefont
  {{Dauchot}}}]{briand2018spontaneously}%
  \BibitemOpen
  \bibfield  {author} {\bibinfo {author} {\bibfnamefont {G.}~\bibnamefont
  {{Briand}}}, \bibinfo {author} {\bibfnamefont {M.}~\bibnamefont
  {{Schindler}}},\ and\ \bibinfo {author} {\bibfnamefont {O.}~\bibnamefont
  {{Dauchot}}},\ }\href {https://doi.org/10.1103/PhysRevLett.120.208001}
  {\bibfield  {journal} {\bibinfo  {journal} {\prl}\ }\textbf {\bibinfo
  {volume} {120}},\ \bibinfo {eid} {208001} (\bibinfo {year} {2018})},\ \Eprint
  {https://arxiv.org/abs/1709.03844} {arXiv:1709.03844 [cond-mat.soft]}
  \BibitemShut {NoStop}%
\bibitem [{\citenamefont {van~der Linden}\ \emph {et~al.}(2019)\citenamefont
  {van~der Linden}, \citenamefont {Alexander}, \citenamefont {Aarts},\ and\
  \citenamefont {Dauchot}}]{vanderlinden2019interrupted}%
  \BibitemOpen
  \bibfield  {author} {\bibinfo {author} {\bibfnamefont {M.~N.}\ \bibnamefont
  {van~der Linden}}, \bibinfo {author} {\bibfnamefont {L.~C.}\ \bibnamefont
  {Alexander}}, \bibinfo {author} {\bibfnamefont {D.~G. A.~L.}\ \bibnamefont
  {Aarts}},\ and\ \bibinfo {author} {\bibfnamefont {O.}~\bibnamefont
  {Dauchot}},\ }\href {https://doi.org/10.1103/PhysRevLett.123.098001}
  {\bibfield  {journal} {\bibinfo  {journal} {Phys. Rev. Lett.}\ }\textbf
  {\bibinfo {volume} {123}},\ \bibinfo {pages} {098001} (\bibinfo {year}
  {2019})}\BibitemShut {NoStop}%
\bibitem [{\citenamefont {{Maitra}}\ and\ \citenamefont
  {{Ramaswamy}}(2019)}]{maitra2019oriented}%
  \BibitemOpen
  \bibfield  {author} {\bibinfo {author} {\bibfnamefont {A.}~\bibnamefont
  {{Maitra}}}\ and\ \bibinfo {author} {\bibfnamefont {S.}~\bibnamefont
  {{Ramaswamy}}},\ }\href {https://doi.org/10.1103/PhysRevLett.123.238001}
  {\bibfield  {journal} {\bibinfo  {journal} {\prl}\ }\textbf {\bibinfo
  {volume} {123}},\ \bibinfo {eid} {238001} (\bibinfo {year} {2019})},\ \Eprint
  {https://arxiv.org/abs/1812.01374} {arXiv:1812.01374 [cond-mat.soft]}
  \BibitemShut {NoStop}%
\bibitem [{SUP()}]{SUPP}%
  \BibitemOpen
  \href@noop {} {\bibinfo  {journal} {{See Supplementary Material at
  [to-be-inserted-by-publisher] for...}}\ }\BibitemShut {NoStop}%
\bibitem [{\citenamefont {Engel}\ \emph {et~al.}(2013)\citenamefont {Engel},
  \citenamefont {Anderson}, \citenamefont {Glotzer}, \citenamefont {Isobe},
  \citenamefont {Bernard},\ and\ \citenamefont {Krauth}}]{engel2013hard}%
  \BibitemOpen
\bibfield  {journal} {  }\bibfield  {author} {\bibinfo {author} {\bibfnamefont
  {M.}~\bibnamefont {Engel}}, \bibinfo {author} {\bibfnamefont {J.~A.}\
  \bibnamefont {Anderson}}, \bibinfo {author} {\bibfnamefont {S.~C.}\
  \bibnamefont {Glotzer}}, \bibinfo {author} {\bibfnamefont {M.}~\bibnamefont
  {Isobe}}, \bibinfo {author} {\bibfnamefont {E.~P.}\ \bibnamefont {Bernard}},\
  and\ \bibinfo {author} {\bibfnamefont {W.}~\bibnamefont {Krauth}},\ }\href
  {https://doi.org/10.1103/PhysRevE.87.042134} {\bibfield  {journal} {\bibinfo
  {journal} {Phys. Rev. E}\ }\textbf {\bibinfo {volume} {87}},\ \bibinfo
  {pages} {042134} (\bibinfo {year} {2013})}\BibitemShut {NoStop}%
\bibitem [{Note1()}]{Note1}%
  \BibitemOpen
  \bibinfo {note} {Measurements performed using the more traditional, but
  oscillatory correlation function $g(r) = \langle \sum _{j\ne k}\delta
  (r-|{\bf r}_j - {\bf r}_k|) e^{i{\bf G}\cdot [{\bf r}_j - {\bf r}_k]}/\sum
  _{j\ne k}\delta (r-|{\bf r}_j - {\bf r}_k|) \rangle _t$, shown in \cite
  {SUPP}, give essentially the same results, albeit with not as clean an
  algebraic decay as observed with $\protect \hat {g}(r)$.}\BibitemShut {Stop}%
\bibitem [{\citenamefont {Kardar}(2007)}]{kardar_2007}%
  \BibitemOpen
  \bibfield  {author} {\bibinfo {author} {\bibfnamefont {M.}~\bibnamefont
  {Kardar}},\ }\href {https://doi.org/10.1017/CBO9780511815881} {\emph
  {\bibinfo {title} {Statistical Physics of Fields}}}\ (\bibinfo  {publisher}
  {Cambridge University Press},\ \bibinfo {year} {2007})\BibitemShut {NoStop}%
\bibitem [{Note2()}]{Note2}%
  \BibitemOpen
  \bibinfo {note} {Note that $\mu $ and $\lambda $ in principle depend on
  temperature, something we neglect in this rough argument.}\BibitemShut
  {Stop}%
\bibitem [{Note3()}]{Note3}%
  \BibitemOpen
  \bibinfo {note} {For $\kappa =0$ (no alignment), unsurprisingly, the spatial
  spectrum is flat. This does not necessarily mean that $T_S = T_D$, since the
  ${\protect \bf s}$ field is still time-correlated}\BibitemShut {NoStop}%
\bibitem [{\citenamefont {Chaikin}\ and\ \citenamefont
  {Lubensky}(1995)}]{chaikin_lubensky_1995}%
  \BibitemOpen
  \bibfield  {author} {\bibinfo {author} {\bibfnamefont {P.~M.}\ \bibnamefont
  {Chaikin}}\ and\ \bibinfo {author} {\bibfnamefont {T.~C.}\ \bibnamefont
  {Lubensky}},\ }\href {https://doi.org/10.1017/CBO9780511813467} {\emph
  {\bibinfo {title} {Principles of Condensed Matter Physics}}}\ (\bibinfo
  {publisher} {Cambridge University Press},\ \bibinfo {year}
  {1995})\BibitemShut {NoStop}%
\bibitem [{Note4()}]{Note4}%
  \BibitemOpen
  \bibinfo {note} {See, e.g., discussion of the time-dependent Ginzburg-Landau
  equation in Kardar's book \protect \cite {kardar_2007}.}\BibitemShut {Stop}%
\bibitem [{Note5()}]{Note5}%
  \BibitemOpen
  \bibinfo {note} {{In the $\kappa =0$ non-aligning limit, where the {\protect
  \bf s} spectrum is flat, one has $a=D_r$, and thus $T_S=\protect \genfrac
  {}{}{}1{1}{2}s_0^2/D_r$, as expected for active Brownian particles, see e.g.
  \cite {cates2015motility}.}}\BibitemShut {Stop}%
\bibitem [{Note6()}]{Note6}%
  \BibitemOpen
  \bibinfo {note} {Note that this is only observed beyond a crossover scale of
  the order of 100 lattice steps, a sizeable numerical difficulty, which comes
  in addition to having to use small timesteps.}\BibitemShut {Stop}%
\bibitem [{Note7()}]{Note7}%
  \BibitemOpen
  \bibinfo {note} {The reader might wonder why most of our results are for the
  rather 'exotic' case of weak alignment. The main reason is practical: one can
  use much larger time-steps in this case than with the hardcore WCA potential,
  saving about one order of magnitude of computation time.}\BibitemShut {Stop}%
\bibitem [{\citenamefont {Cates}\ and\ \citenamefont
  {Tailleur}(2015)}]{cates2015motility}%
  \BibitemOpen
  \bibfield  {author} {\bibinfo {author} {\bibfnamefont {M.~E.}\ \bibnamefont
  {Cates}}\ and\ \bibinfo {author} {\bibfnamefont {J.}~\bibnamefont
  {Tailleur}},\ }\href
  {http://www.annualreviews.org/doi/abs/10.1146/annurev-conmatphys-031214-014710}
  {\bibfield  {journal} {\bibinfo  {journal} {Annu. Rev. Condens. Matter
  Phys.}\ }\textbf {\bibinfo {volume} {6}},\ \bibinfo {pages} {219} (\bibinfo
  {year} {2015})}\BibitemShut {NoStop}%
\end{thebibliography}%

\end{document}